\documentclass[fleqn,10pt]{wlscirep}
\usepackage[utf8]{inputenc}
\usepackage[T1]{fontenc}
\usepackage{subfigure}
\usepackage{xcolor} 
\usepackage{pgfplots}
\title{Low dimensional behaviour of generalized Kuramoto model} 

\author[1,*]{Sara Ameli}
\author[2]{Keivan Aghababaei Samani}

\affil[1]{Max Planck Institute for Physics of Complex Systems, Dresden  01187,  Germany.}
\affil[2]{Department of Physics, Isfahan University of Technology, Isfahan 84156-83111, Iran.}
\affil[*]{sameli@pks.mpg.de}

\begin{abstract}
We study the global bifurcations of frequency weighted Kuramoto model in low-dimension for network of fully connected oscillators. To study the effect of non-zero-centered frequency distribution, we consider two symmetric Lorentzians as an example. 
We derive the stability diagram of the system and show that the infinite-dimensional problem reduces to a flow in four dimensions. Using the system symmetries, it can be further reduced to two dimensions. Using this analytic framework, we obtain bifurcation boundaries of the system, which is compatible with our numeric simulations. We show that
the system has three types of transitions to synchronized state for different parameters of the frequency distribution: (1) a two-step transition, representative of standing waves, (2) a continuous transition, as in the classical Kuramoto model, and (3) a
first-order transition with hysteresis. Numerical simulations are also conducted to confirm analytic results.
\end{abstract}

\begin{document}
\flushbottom
\maketitle

\thispagestyle{empty}

\section{\label{sec:level1}Introduction}
Synchronization is a phenomenon representing the collective behavior in populations of interacting units.  
Examples are the synchronous flashing of groups of fireflies \cite{buck1938synchronous}, oscillation of neuronal communication causing circadian rhythms \cite{herzog2007neurons}, chemically reacting cells \cite{kuramoto2003chemical}, Josephson junction circuits \cite{wiesenfeld1996synchronization}, etc. A key contribution in this area was introduced by Kuramoto \cite{kuramoto1975self}$^, $ \cite{rodrigues2016kuramoto} . The Kuramoto model describes the system's transition to collective dynamics with the increase in coupling strength bypassing a threshold. 
The original Kuramoto model considers a solid interaction between dynamical units. Even though it describes the collective phenomena, it misses the interplay between structure and dynamics.
However, the generalized model, inspired by natural systems, considers the correlation between structure and dynamics by considering connections as a function of the intrinsic dynamics of units. Properties of transition from disordered (asynchronous) to ordered (synchronous) state are of particular interest. Synchronization is commonly found to follow a smooth, second-order phase transition, as shown generally by Kuramoto.
However, it has been shown that in some cases, there is a first-order, or discontinuous synchronization transition,
i.e., an abrupt and irreversible phase transition to the synchronized
state, which is called {\it explosive synchronization (ES)}. The first ES was reported in scale-free network topology with a positive correlation between natural frequencies of the oscillators and their degrees\cite{gomez2011explosive}. It has been shown that even if there is a partial correlation between natural frequencies of nodes and their degrees, the transition would be first-order \cite{pinto2015explosive}.
This type of transition has been studied from different perspectives in \cite{ peron2012explosive, skardal2014disorder, leyva2012explosive, kaue2012determining, pazo2005thermodynamic, zhang2015explosive, huang2016effects, ji2013cluster, ji2014analysis, li2013reexamination, zhu2013criterion, sendina2015effects, chen2015self}.
Recently, experimental evidence of explosive synchronization in electrical, chemical, and neuronal systems are reported~\cite{kumar2015experimental, leyva2012explosive, chen2013explosive}. The first-order transition causes a hysteretic behavior in the system. The origin of hysteretic behavior has been investigated in \cite{zou2014basin, torcini2015hysteretic}. In particular, ES can also happen when there is a positive correlation between the coupling strengths of the oscillators and the absolute value of the difference in their natural frequencies \cite{zhang2013explosive}. In \cite{hu2014exact} they have obtained the exact value of the critical coupling for transition to the synchronized state. 
To study the transition of the system to the synchronized state, we use the Ott-Antonsen ansatz \cite{ott2008low} to lower the dimension of the infinite-dimensional system. In this way, one can investigate the effect of frequency distribution on the evolution of the order parameter and phase transition as a function of the coupling strength.


\section{\label{sec:level2}The frequency weighted Kuramoto model}

The Kuramoto model with interactions weighted by the absolute value of frequencies, considers an ensemble of $N$ coupled limit-cycle oscillators. The phase of $i$'th oscillator is given by  $\theta_{i}$, and its evolution is governed by
\begin{equation}
\frac{d\theta_{i}}{dt}=\omega_{i}+\frac{\lambda \vert \omega_{i}\vert}{k_{i}} \sum_{j=1} ^{N} A_{ij} \sin({\theta _j-\theta_i}) ,~~      i=1,2,...,N.
\label{kuramoto2}
\end{equation} 
Where $\lbrace\omega_{i}\rbrace$, the natural frequency of the $i$th oscillator, is drown from a frequency distribution. The connectivity pattern is given by $A_{ij}$s that are the elements of adjacency matrix, representing the topology of interacting network, with  $A_{ij}=1$ for connected nodes and $A_{ij}=0$ otherwise. Therefore, $k_{i}=\sum_{j}A_{ij}$ is the degree of node $i$. $\lambda$ stands for the strength of coupling. The complex order parameter, as a measure of synchrony, can be defined as:
\begin{equation}
    r=\int_{0}^{2\pi}e^{i\theta}\Big(\frac{1}{N}\sum_{j}^{N} \delta(\theta-\theta_j)\Big)d\theta
    \label{rd}
\end{equation}
The significant characteristic of this model is the frequency-weighted coupling, which causes a positive correlation between the coupling strength of oscillators and the absolute value of their natural frequencies, providing a heterogeneous interaction between units. This kind of coupling is characteristic of some natural and synthesis systems, such as power grids and social interactions. For example, introverted people less frequently interact with the rest of society. Hence, the rate of adaptation is lower for them.

\subsection{\label{sec:level3}Continuum limit}

In the limit of the infinite number of oscillators $(N\rightarrow \infty)$, for each natural frequency $\omega$, there is a continuum of oscillators distributed around the unit circle. 
We define this distribution with $\rho(\theta,\omega,t)$, so that $\rho(\theta,\omega,t) d\theta$ is the fraction of oscillators of natural frequency $\omega$, whose phase is between $\theta$ and $\theta+d\theta$ at time $t$. $\rho(\theta,\omega,t)$ is a $2\pi$ periodic function and follows the normalization condition:
\begin{equation}
\int_{0}^{2\pi}\rho(\theta,\omega,t)d\theta=1 
\label{rhoN}
\end{equation}
and frequency distribution also is supposed to be normalized:
\begin{equation}
\int_{-\infty}^{\infty}g(\omega)d\omega=1 
\label{g}
\end{equation}
Considering equations \ref{rhoN} and \ref{g} we have:
\begin{equation}
\int_{-\infty}^{\infty} \int_{0}^{2\pi}\rho(\theta,\omega,t)g(\omega)d\theta d\omega=1 
\end{equation}
Having conservation of oscillators of frequency $\omega$, the evolution of $\rho$ is governed by the continuity equation:
\begin{equation}
\frac{\partial \rho}{\partial t}+\frac{\partial(\rho v)}{\partial \theta}=0
\label{rho}
\end{equation}
The velocity in this continuous representation can be stated as:
\begin{equation}
\small
v(\theta, \omega, t)=\omega+\lambda|\omega|\int_{-\infty}^{\infty}\int_{0}^{2\pi} \rho(\theta',\omega,t)g(\omega)\sin(\theta'-\theta) d\theta' d\omega
\label{V}
\end{equation}
or alternatively as:
\begin{equation}
v(\theta, \omega,t)=\omega+\lambda\vert \omega\vert Im(re^{-i\theta}) \\
\end{equation}
with $r$, complex order parameter, defined with the following non-linear integro- partial-differential equation:
\begin{equation}
r(t)=\int_{0}^{2\pi}\int_{-\infty}^{+\infty}e^{i\theta'}\rho(\theta', \omega, t) g(\omega) d\theta' d\omega
\label{rcon}
\end{equation}
Which is the continuous version of Eq.(\ref{rd}). In the incoherent state, which corresponds to the case of small coupling strength, the oscillators are distributed around unit circle $\rho_{0}(\theta, \omega, t)=\frac{1}{2\pi}$, having their own intrinsic dynamic. By finding the point that this solution loses it's stability, one can find the critical coupling in which the system does transition to synchronized state, which is the point that the system looses it's incoherent state. To do so, we add a perturbation $(\eta)$ to incoherent density of oscillators:
\begin{equation}
\rho(\theta,\omega,t)=\frac{1}{2\pi}+\epsilon\eta(\theta, \omega, t)
\label{ro}
\end{equation}
where $\epsilon\ll 1$, and $\eta$ is real, $2\pi$-periodic, and the normalization condition Eq. (\ref{rhoN}) requires that $\int_{0}^{2\pi}\eta(\theta, \omega, t)d\theta=0$.

We restrict ourselves to linear analysis about the fixed point and we expand $\eta$ in it's Fourier modes $\eta(\theta, \omega, t)=\sum_n\eta_n(\omega, t) e^{i n\theta}$  and $\eta_{n}(\omega,t)=\alpha(\omega,t)^n$, as considered by Ott-Antonsen \cite{ott2008low}. $|\alpha(\omega, t)|\leq 1$ to avoid divergence of series.
This ansatz provides a family of distribution functions that provide information about the synchronized and non-synchronized dynamic naturally. By plugging this ansatz into Eq. (\ref{ro}), the density can be written as:
\begin{equation}
\rho(\theta, \omega,t)=\frac{1}{2\pi}\Big[1+\sum_{n=1}^{\infty}(\alpha(\omega, t) e^{i\theta})^n+\sum_{n=1}^{\infty}(\alpha^{*}(\omega, t)e^{-i\theta})^n\Big]\\
\end{equation}
Therefore, the continuity equation (Eq. (\ref{rho})) becomes:
\begin{equation}
\small
\begin{split}
&\frac{\partial}{\partial t}\Big[1+\sum_{n=1}^{\infty}(\alpha(\omega, t) e^{i\theta})^n+\sum_{n=1}^{\infty}(\alpha^{*}(\omega, t)e^{-i\theta})^n\Big]+\\
&\frac{\partial}{\partial\theta}\Bigg[\Big[1+\sum_{n=1}^{\infty}(\alpha(\omega, t) e^{i\theta})^n+\sum_{n=1}^{\infty}(\alpha^{*}(\omega, t)e^{-i\theta})^n\Big]\\
&~~~~~~~~(\omega+\frac{\lambda}{2i} \vert \omega\vert (re^{-i\theta}-r^*e^{i\theta}))\Bigg]=0
\end{split}
\end{equation}

Since different frequency modes evolve independently, one can study them separately. By considering the lower order terms in the sum in both sides of the equation, we obtain:
\begin{equation}
\frac{\partial \alpha}{\partial t}+   \frac{\lambda\vert \omega\vert}{2}(-r^{*}+\alpha^{2}r)  + i \alpha \omega  =0
\end{equation}
and
\begin{equation}
\frac{d\alpha^*}{dt}+\frac{\lambda\vert \omega\vert }{2}(-r+\alpha^{*2}r^*)-i\alpha^*\omega=0
\label{mainalpha}
\end{equation}
For frequency distribution, we consider bimodal Lorentzian distribution with four poles at $\omega=\pm \omega_{0}\pm i\Delta$. The distribution in partial fraction form, can be written as:
\begin{equation}
\begin{split}
&g(\omega)=\frac{1}{4\pi i}\Big[  \frac{1}{(\omega-\omega_{0})-i\Delta} -\frac{1}
{(\omega-\omega_{0})+i\Delta}\\
&~~~~~~~~~~~~~+\frac{1}{(\omega+\omega_{0})-i\Delta}    -\frac{1}{(\omega+\omega_{0})+i\Delta}           \Big]
\end{split}
\end{equation}

\begin{figure*}[t!]
\centering
    \includegraphics[scale=0.45]{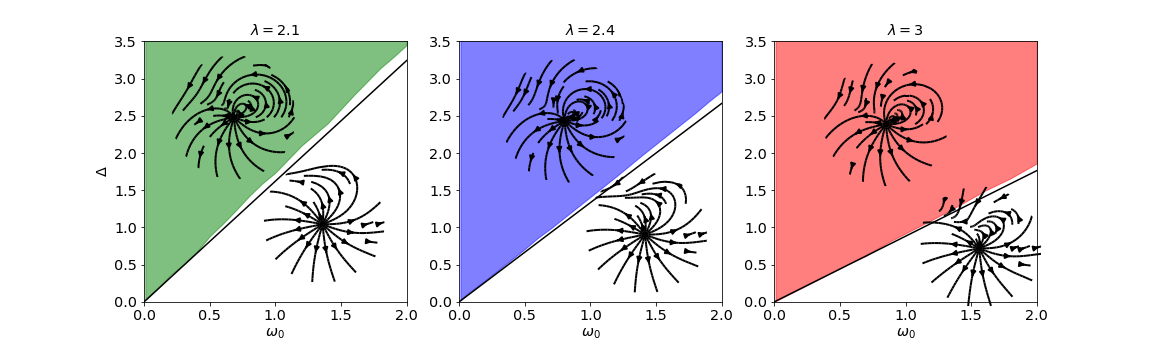}
    \caption{Flow field around zero, based on the Eq. (\ref{qsi}) for $\omega_0=2$, together with the results of the linear stability analysis Eq. (\ref{B}) shown with black solid lines. Colors are representative of three different value for coupling strength ($\lambda)$.}
\label{FF}    
\end{figure*}

To make sure that $g(\omega)$ is concave in the center and therefore can be considered bimodal, it has to fulfill the condition $\omega_0/\Delta >1/\sqrt{3}$. Otherwise, the tails of the two distributions merge and they form a unimodal distribution. 
 We consider $\alpha(\omega,t)$ to be continuous in the complex $\omega$ plane. Evaluating $r^*=\int _{-\infty}^{\infty}\alpha(\omega,t)g(\omega)d\omega$, by deforming the integration from the real $\omega $ axis to $Im(\omega)\longrightarrow-\infty$, the order parameter becomes:
$r=\frac{r_1+r_2}{2}$, where $r_{1,2}(t)=\alpha^{*}(\pm\omega_{0}-i\Delta, t)$. By plugging these values into Eq. (\ref{mainalpha}),
the system reduces to a set of two coupled complex ODEs for two sub order parameters:
\begin{equation}
\small
\begin{split}
&\text{using}  ~\alpha^*=r_1~,~~~\omega=-\omega_0-i\Delta, ~\text{we have:}\\
&\frac{dr_1}{dt}=\frac{\lambda\vert-\omega_0-i\Delta\vert}{4}\big(r_1+r_2-r_1^2(r^*_1+r^*_2)\big)+ir_1(-\omega_0-i\Delta)\\
&\text{and for}~\alpha^*=r_2~,~~~\omega=~~~\omega_0-i\Delta,~\text{we get:}\\
&\frac{d r_2}{dt}=\frac{\lambda\vert~~\omega_0-i\Delta\vert }{4}\big(r_1+r_2-r_2^2(r_1^*+r_2^*)\big)+ir_2(~~\omega_0-i\Delta)
\label{r1r2}
\end{split}
\end{equation}
Which provides information about the dynamics of the two clusters in the system. Since $\alpha$ is the first mode of perturbation when it is non-zero, it shows deviation from the random state.

\subsection{Linear stability}

We analyse the linear stability of the incoherent state by perturbing the system around the fixed points. By adding perturbations ( $r_1 \rightarrow r_1 +dr_1$ and $r_2 \rightarrow r_2 +dr_2$) to Eq. (\ref{r1r2}),

and keeping the linear terms we have:
\begin{equation}
\begin{split}
&\dot{r}_1= \frac{\lambda\vert-\omega_0-i\Delta\vert}{4}[r_1+r_2]  - r_1(i\omega_0-\Delta)\\
&\dot{r}_2=   \frac{\lambda \vert\omega_0-i\Delta\vert}{4}~~[r_1+r_2]  -  r_2(-i\omega_0-\Delta)\\
\label{r1dotr2dot}
\end{split}
\end{equation}
Which can be written in a matrix form:
\begin{equation}
\small
\begin{pmatrix}
\dot{r}_1 \\
\dot{r}_2 
\end{pmatrix}=
\begin{pmatrix}
\frac{\lambda\vert-\omega_0-i\Delta\vert}{4}-(i\omega_0-\Delta) & \frac{\lambda\vert-\omega_0-i\Delta\vert}{4}\\
\frac{\lambda\vert\omega_0-i\Delta\vert}{4} & \frac{\lambda\vert\omega_0-i\Delta\vert}{4}-(-i\omega_0-\Delta)
\end{pmatrix}
\begin{pmatrix}
{r}_1 \\
{r}_2 
\end{pmatrix}
\end{equation}
The eigenvalues of this matrix ($\Lambda_1,\Lambda_2$) can be found:
\begin{equation}
\Lambda_{1,2}=\frac{\lambda\sqrt{\omega_0^2+\Delta^2}}{4}+\Delta\pm\sqrt{\frac{\lambda^2(\omega_0^2+\Delta^2)}{16}-\omega_0^2}
\end{equation}
For the second square root to be real, it has to fulfill the condition:
\begin{equation}
    \frac{\lambda^2(\omega_0^2+\Delta^2)}{16}-\omega_0^2 \geqslant 0
    \label{Delt}
\end{equation}
Which indicates the bifurcation boundary of the system at 
\begin{equation}
    \Delta = \omega_0 \sqrt{\frac{16-\lambda^2}{\lambda^2}}
    \label{B}
\end{equation}   
Shown with solid black lines in FIG. (\ref{FF}). 
For $
    \Delta \geqslant \omega_0 \sqrt{\frac{16-\lambda^2}{\lambda^2}}~   \text{or}~  \Delta \leqslant  -\omega_0 \sqrt{\frac{16-\lambda^2}{\lambda^2}}
    \label{DeltS}
$, both eigenvalues are real. Otherwise, they have imaginary parts as well. Since it is symmetric around zero, we just plot the upper half plane. In the next section we will study the dynamic of the flow field in these two regions.

\subsection{Reduction of the system  }
We define complex order parameters in polar coordinate, and a phase difference between two sub-order parameters as:
\begin{equation}
r_1=\rho_1 e^{i\phi_1}~,~~~~~~~r_2=\rho_2 e^{i\phi_2}~,~~~~~~~\psi=\phi_2-\phi_1
\end{equation}
Plugging in Eqs. (\ref{r1r2}) and comparing real parts in both sides of the equation, we get a differential equation for $\rho$, the radius of the deviation from the random state, and by comparing imaginary parts, we get a differential equation for the evolution of phase $(\phi)$ of each cluster.
Using the expressions obtained for $\dot{\phi_2}$ and $\dot{\phi_1}$ we can find an expression for $\dot{\psi}$.
So, we have a set of three differential equations for the amplitude and phase difference of suborder parameters.
\begin{equation}
\begin{split}
&\dot{\rho}_1~= \frac{\lambda \sqrt{\omega_0^2+\Delta^2}}{4}\big[(1-\rho_1^2)(\rho_1+\rho_2\cos\psi)\big] + \rho_1 \Delta\\
&\dot{\rho}_2~=\frac{\lambda\sqrt{\omega_{0}^2+\Delta^2}}{4}\big[(1-\rho_2^2)(\rho_1\cos\psi+\rho_2)\big] +  \rho_2 \Delta\\
&\dot{\psi}=-\frac{\lambda\sqrt{\omega_0^2+\Delta^2}}{4}\big[\frac{\rho_1^2+\rho_2^2+2\rho_1^2\rho_2^2}{\rho_1\rho_2}\big]\sin\psi+2\omega_0\\
\end{split}
\end{equation}
We look for solutions that satisfy the symmetry condition $\rho_1(t)=\rho_2(t)=\rho(t)$. Considering this condition, the dimension of the system reduces to two.
\begin{equation}
\begin{split}
&\dot{\rho}=\frac{\lambda \sqrt{ \omega_0^2+\Delta^2}}{4}\big[(1-\rho^2)(\rho+\rho\cos\psi)\big] + \rho \Delta\\
&\dot{\psi}=-\frac{\lambda\sqrt{\omega_0^2+\Delta^2}}{4}\big[\frac{\rho^2+\rho^2+2\rho^2\rho^2}{\rho\rho}\big]\sin\psi+2\omega_0\\
\end{split}
\end{equation}
By defining: $q=\rho^2$ then $\dot{q}=2\dot{\rho}\rho$, and multiplying both sides of equation for $\dot{\rho}$ with $2\rho$
The first equation becomes:
\begin{equation}
\begin{split}
&2\rho\dot{\rho}=\rho^2\frac{\lambda \sqrt{ \omega_0^2+\Delta^2}}{2}\big[(1-\rho^2)(1+\cos\psi)\big] + 2\rho^2 \Delta\\
\end{split}
\end{equation}
We end up with a two dimensional phase plane:
\begin{equation}
\begin{split}
&\dot{q}~=q\frac{\lambda\sqrt{\omega_0^2+\Delta^2}}{2}~~~\big[(1-q)~+~(1-q)\cos\psi\big]+  2q \Delta\\
&\dot{\psi}=-\frac{\lambda\sqrt{\omega_0^2+\Delta^2}}{2}\big[1+q\big]\sin\psi+2\omega_0\\
\label{qsi}
\end{split}
\end{equation}

We use equations in (\ref{qsi}) to plot dynamical flow in the $q-\psi$ plane. In FIG. \ref{FF}, the colored region shows the parameter values $(\Delta, \omega_0)$, which have the same type of dynamic, same for the white region. In this case, zero is a saddle node in the colored regions and, in the white regions, an unstable fixed point. In shadowed regions, both $\Lambda_1$ and $\Lambda_2$ are real, and zero is a saddle node. While, in the white regions, they have imaginary parts as well, leading to an unstable fixed point at zero. Different colors represent different values of $\lambda$. The color green stands for $\lambda=2.1$, the color blue for the case of $\lambda=2.4$, and red for $\lambda=3$. There is a good match between the solid line, obtained from Eq.(\ref{B}), and the boundary of the shaded region obtained from flow field characteristic using Eq.(\ref{qsi}). The little mismatch is due to the higher-order terms, which we have ignored.

\subsection{Bifurcation analysis}

Using set of equations in (\ref{qsi}), the $q=0$ case, corresponding to the incoherent state, gives rise to $\dot{q}=0~,\dot{\psi}=-\frac{\lambda\sqrt{\omega_0^2+\Delta^2}}{2}\sin\psi+2\omega_0$. If in addition to $q=0$, $\dot{q}=0$, $\dot{\psi}$ is also equal to zero, then we have $\psi=\sin^{-1}\Big[\frac{4\omega_0}{\lambda\sqrt{\omega_0^2+\Delta^2}}\Big]$. 
In addition to the trivial incoherent state q=0, the other fixed points of the system occurs when $q\neq 0$, $\dot{q}=0$, $\dot{\psi}=0$. putting these conditions in Eq.(\ref{qsi}), we can find two trigonometric equations:
\begin{equation}
\begin{split}
&\cos\psi=\big[\frac{-4\Delta}{\lambda\sqrt{\omega_0^2+\Delta^2}}-(1-q)\big]\frac{1}{1-q}\\
&\sin\psi=\frac{4\omega_0}{\lambda\big[1+q\big]\sqrt{\omega_0^2+\Delta^2}}
\end{split}
\end{equation}
Using the trigonometric relation $\sin^2+\cos^2=1$, one obtains a third order polynomial equation for fixed points of q:

\begin{equation}
\begin{split}
&q^3(-\frac{\Delta\lambda\sqrt{\omega_0^2+\Delta^2}}{2})+(\omega_0^2+\Delta^2-\frac{\Delta\lambda\sqrt{\omega_0^2+\Delta^2}}{2})q^2\\
&+\big((-\omega_0^2+\Delta^2)2+\frac{\Delta\lambda\sqrt{\omega_0^2+\Delta^2}}{2}\big)q+\omega_0^2+\Delta^2\\
&+\frac{\Delta\lambda\sqrt{\omega_0^2+\Delta^2}}{2} =0
\end{split}
\end{equation}
Using the characteristic equation of the third order polynomial $\delta = 18abcd-4b^3 d+b^2 c^2-4ac^3-27a^2 d^2 $, which gives information about the roots of the polynomial, we can find the bifurcation in the system.
\begin{equation}
\begin{split}
&\delta=4 (-\frac{\Delta\lambda\sqrt{\omega_0^2+\Delta^2}}{2}) (\omega_0^2+\Delta^2)\big[-(\omega_0^2+\Delta^2)^2\\
&+12(-\frac{\Delta\lambda\sqrt{\omega_0^2+\Delta^2}}{2})^2\big]
\end{split}
\end{equation}
which changes sign when:
\begin{equation}
    -(\omega_0^2+\Delta^2)^2+3\Delta^2\lambda^2(\omega_0^2+\Delta^2)=0
\end{equation}
and simplifies to:
\begin{equation}
    \frac{\omega_0}{\Delta}=\sqrt{3\lambda^2-1}
    \label{w0oD}
\end{equation}
Meaning that $\omega_0/\Delta = \sqrt{3\lambda^2-1}$ is the bifurcation line of the system, and by crossing this line, 
the dynamic of the system changes. 

\section{Numeric approach}
\subsection{Order parameter}

By simulating the generalized Kuramoto model, Eq.~(\ref{kuramoto2}), we investigate the effect of the correlation between coupling constant and intrinsic dynamics of each node in the transition to the synchronized state. We consider a fully connected network. To understand the effect of intrinsic frequencies, we take them from a bimodal Lorentzian distribution; by fixing the {\it mean} of the distributions ($\omega_0=2$), we investigate the impact of the {\it width} of the distributions ($\Delta$) on transition to synchronized state.

\begin{figure*}[!t]
    \centering
    \includegraphics[width=18cm]{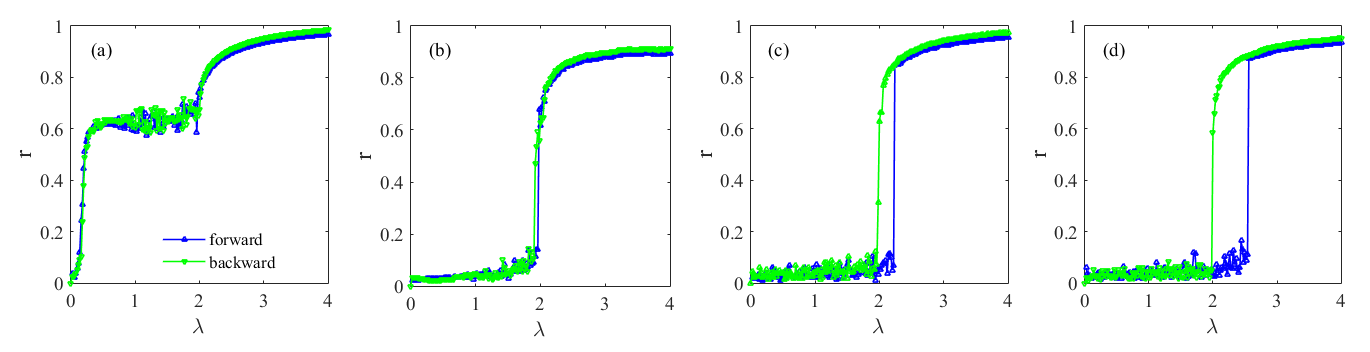}
    \caption{Synchronization in all to all and network with  and bimodal Lorentz distribution for which location of peaks are in 2 and -2,
for different scale parameter, $\Delta$: For (a), $\Delta=0.1$, (b) $\Delta=1$, (c) $\Delta=1.6$, (d) $\Delta=2$.}
    \label{LorentzianR}
\end{figure*}

We compute the stationary value of order parameter $r$, Eq. \ref{rd}, by progressively changing the coupling strength in an adiabatic way and recording the average value at a time window in the steady state. The forward path is obtained by increasing the coupling strength, using the outcome of the previous step as the initial condition. The same progress is performed for backward continuation by decreasing the amount of coupling strength. In all cases, the size of the network is considered to be $N=1000$, and initial phases are taken randomly between $0$ and $2\pi$.
In FIG. \ref{LorentzianR}, we show the results for different values of $\Delta$.  For narrow distributions $(\Delta<1)$, the system has a two-step transition to the synchronized state FIG. \ref{LorentzianR}.(a). In this intermediate step, there are two giant synchronous clusters formed in the system. Inside each cluster, the oscillators are synchronized with each other, but two clusters are not in sync with each other. One of them has a center frequency $\omega_0$ and the other one $-\omega_0$. So they form standing waves in the system. This region remains stable even with increase in the $\lambda$. At the final transition ($\lambda\simeq 2$), these two synchronous clusters join together and make the whole population synchronized.

By simulating for different values of the $\Delta$, we have observed that the smaller the width of the distribution, $\Delta$, the larger this stable region is. By getting closer to $\Delta=1$, the size of this intermediate region shrinks, and close to $\Delta=1$, the system shows a continuous transition to the synchronized state, FIG. \ref{LorentzianR}.(b), similar to the classical Kuramoto model. The reason for such a behaviour is that for small $\Delta$, two clusters are far apart from each other and focused around their natural frequencies $\omega_0$ and $-\omega_0$. The narrower the distribution, the larger coupling is required to combine both clusters. By further increasing the width of the distribution, we observe a first-order transition with hysteresis to the synchronized state for $\Delta>1$ (FIG. \ref{LorentzianR}.(c)). The existence of this hysteresis is indicative of memory in the system. Having a first-order transition to the synchronized state and a hysteresis loop is representation of the {\it explosive synchronization}. Increasing the width of the distribution $\Delta$, causes the forward transition to happen in larger $\lambda$, and causing the system to have a larger bistable region FIG. \ref{LorentzianR}.(d). When there is a hysteretic transition $(\Delta>1)$, the backward transition, shown with green, is always happening at $\lambda^*=2$. While the forward transition, in blue, happens in larger $\lambda$ for wider distribution. The reason is that in the backward transition, the dynamic starts from a synchronized state, so the width of the distribution does not play a role.

\begin{figure*}[!ht]
\centering
\includegraphics[width=14.7cm]{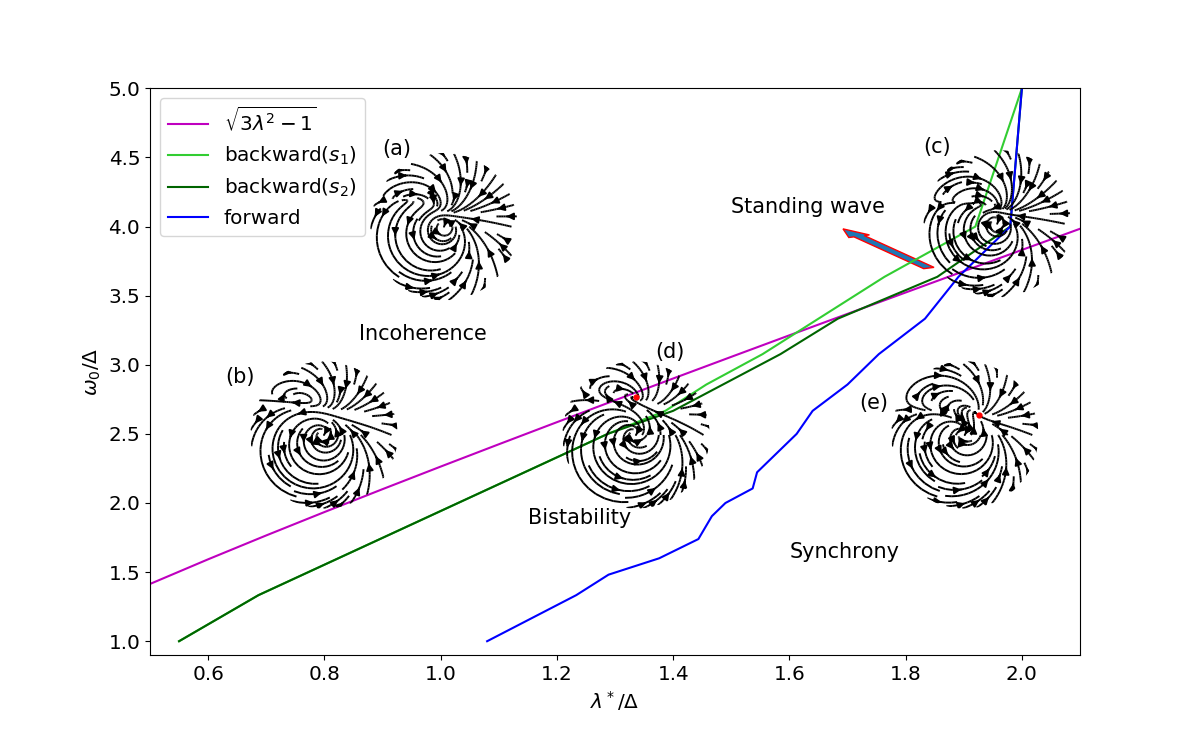}
\caption{Stability diagram. The blue line is the critical value of the forward transition to the synchronized state. The green line shows the 
critical point of the backward path. Both obtained from numerical simulation. The region between the blue and green lines is the bistable region.
The magenta line is the bifurcation line which is obtained from Eq. (\ref{w0oD}). For all the cases $\omega_0=2$. For (a), (c) $\Delta=0.5$, and $\lambda=0.5$, $0.97$ respectively. For (b),(d),(e)  $\Delta=0.8$, $\lambda=0.64$, $1.05$, $1.52$ respectively.}   
\label{Pspace} 
\end{figure*}

In FIG. \ref{Pspace}, we plot the critical value of the transition $(\lambda^*)$ in the parameter space. To do so, we simulate the system for values of $\omega_0$, and $\Delta$; the same procedure as FIG. \ref{LorentzianR}. For simplicity, we fix $\omega_0=2$, and obtain the critical value of the transition ($\lambda^*$) for different values of $\Delta$. Connecting these values gives the solid blue line for the forward critical value and the green lines for the backward transition. The light green shows the initial step of the transition, which leads to the intermediate region. The dark green shows the final step in the transition to the synchronized state. The region between them is the region of the standing wave. In between the blue and green lines, there is a region that corresponds to the hysteresis behavior in which the system shows bistable dynamic for $\Delta>1$.

We use the Eq. (\ref{qsi}) again to plot the flow field in the $q-\psi$ plane for a given value of parameters. The parameters correspond to the value at the center of the circle for each flow field plot. We observe different types of flow field characteristics in different regions of the parameter space. In all cases, zero is an unstable node. In FIG. \ref{Pspace}.(a),(b),(c), there is an unstable non-zero node in the second quarter, which means the synchronized state is unstable. In addition, there exists a stable limit cycle centered at zero. It indicates the fact that there are two giant dynamical clusters around the centers of the distributions. The phase difference between them continues to increase.
In FIG. \ref{Pspace}.(d), there is a fixed point solution which is  representation of the synchronized state. We have observed the same type of dynamic until close to the boundaries of the magenta line. The magenta line obtained from Eq. (\ref{w0oD}), defines the boundaries of the transition. Close to this line, the zero-centered limit cycle disappears, and the fixed point becomes stable (d). Stable fixed point reflects a stable synchronized state. This region corresponds to the bistable dynamic. In the lower part, shown in FIG. \ref{Pspace}.(e), there exists a stable fixed point for which $\dot{q}=0, \dot{\psi}=0$ indicating a stable synchronized state. 



We did simulation for bimodal Gaussian distribution as well and observed similar behavior.

\subsection{ Correlation matrix }

To shed more light on the dynamic of the system, we plot the frequency distribution in the network for two types of dynamic. We select one example from a two-step transition $\Delta=0.8$, and one example from an explosive transition $\Delta=2$. for different values of coupling strength ($\lambda$). Given that frequencies are symmetric, in FIG.\ref{PWF}(a), the overall frequency distribution cancels around zero. By increasing the coupling strength, the distribution becomes asymmetric at (c), which indicates one cluster becoming stronger than the other one. Finally, at (d) $\lambda=2.04$, one cluster takes over. 

In order to study the collective dynamics in more detail, we look at the correlation matrix. The local phase configurations of the system can be determined by the correlation matrix $D$ defined as:
\begin{equation}
D_{ij}= \lim_{\Delta t \to \infty}\frac{1}{\Delta t}\int^{\Delta t+t_s}_{t_s}\cos(\theta_i(t)-\theta_j(t))dt.
\label{D}
\end{equation}
where $t_s$ is the time needed for reaching to stationary state.  The correlation matrix element $D_{ij}$ is a measure of coherency between the pair of oscillators at $i$ and $j$ positions and takes a value in the  interval $1\leq D \leq -1$. $D_{ij}=1$ when there is a full synchrony between nodes $i$ and $j$ ($\theta_i=\theta_j$), and $D_{ij}=-1$ when they are in the anti-phase state ($\theta_i=\theta_j \pm \pi$). Here we used $5\times 10^5$ time steps for the stationary time $t_s$ and $3\times 10^4$  time steps for the averaging window $\Delta t$.

\subsubsection{Correlation matrix, frequency distribution (forward path)}
\begin{figure*}[!ht]
\includegraphics[scale=0.39]{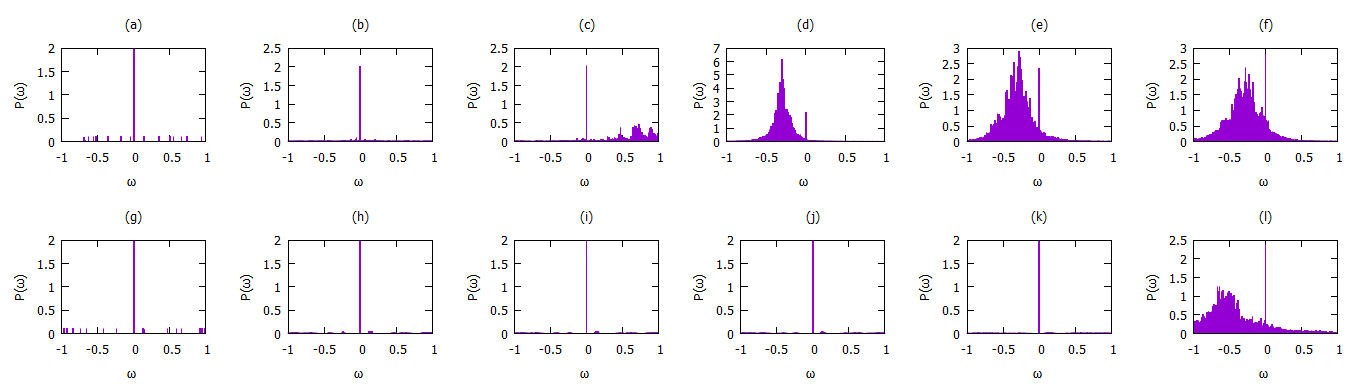}
\caption{Probability density function of the frequency (P($\omega$)) in forward transition. The first row corresponds to the case of $\Delta=0.8$, and second row for the last row $\Delta=2$. Each colum corresponds to the same value of $\lambda$. In (a), (g) $\lambda=0.02$, in (b), (h) $\lambda=1.8$, in (c), (i) $\lambda=1.98$, n (d), (j) $\lambda=2.04$, in (e), (k) $\lambda=2.98$, in (f), (l) $\lambda=3$.)}
\label{PWF}
\end{figure*}

\begin{figure*}[!ht]
\includegraphics[scale=0.39]{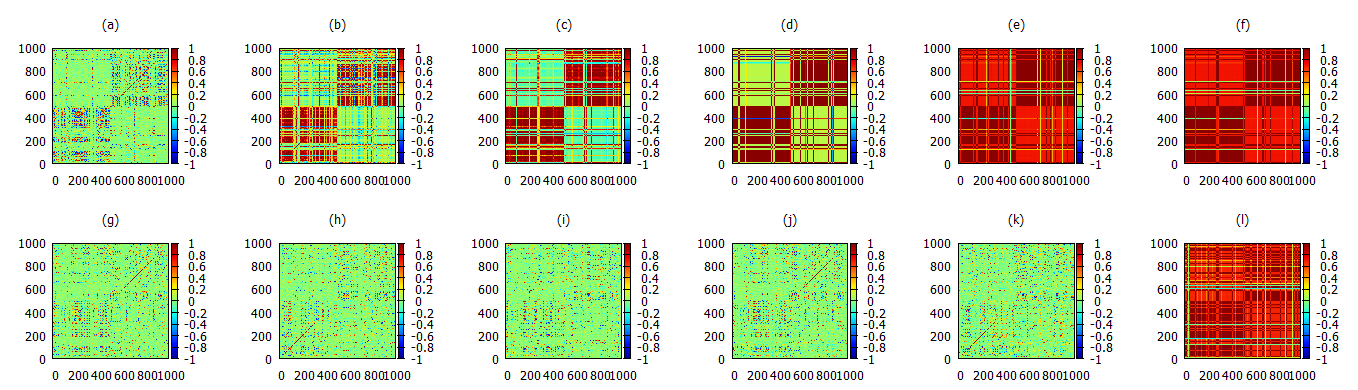}
\caption{Correlation matrix $(D_{ij})$ in forward transition. The first row corresponds to the case of $\Delta=0.8$, and second row for the last row $\Delta=2$. Each column corresponds to the same value of $\lambda$. In (a), (g) $\lambda=0.02$, in (b), (h) $\lambda=1.8$, in (c), (i) $\lambda=1.8$, in (d), (j) $\lambda=1.92$, in (e), (k) $\lambda=2.98$, and in (f), (l) $\lambda=3$.}
\label{SuF}
\end{figure*}

\begin{figure*}[!ht]
\includegraphics[scale=0.39]{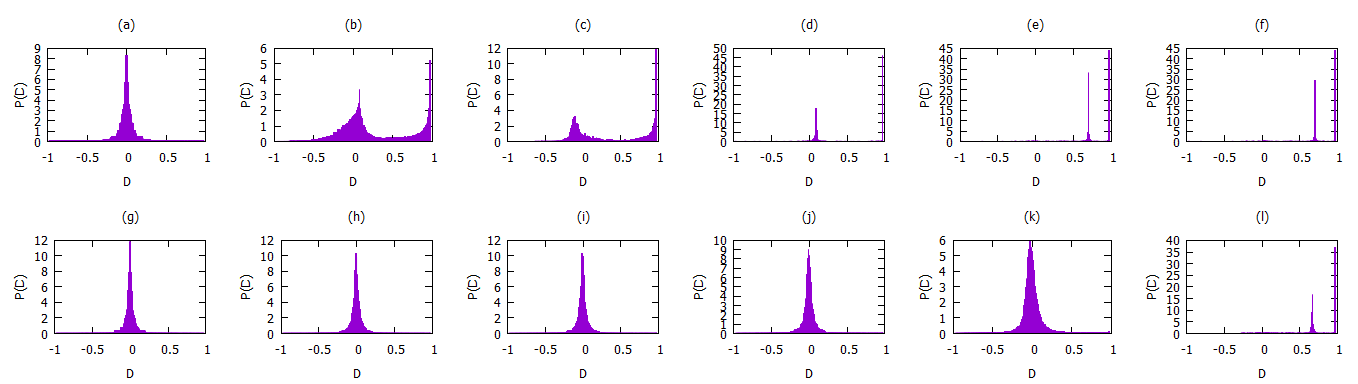}
\caption{Probability distribution of the correlation matrix P$(D_{ij})$ in forward transition. The first row corresponds to the case of $\Delta=0.8$, and second row for the last row $\Delta=2$. Each column corresponds to the same value of $\lambda$. In (a), (g) $\lambda=0.02$, in (b), (h) $\lambda=1.8$, in (c), (i) $\lambda=1.8$, in (d), (j) $\lambda=1.92$, in (e), (k) $\lambda=2.98$, and in (f), (l) $\lambda=3$.}
\label{PDF}
\end{figure*}

We calculate the time-averaged
correlation matrices $(D_ij)$ in FIG. \ref{SuF}, as well as the probability density functions of their elements and represent then in FIG. \ref{PDF}.
The first row shows the results of $\Delta=0.8$, and the second row corresponds to the case of $\Delta=2$. The coupling strength ($\lambda$) are the same in FIGs. (\ref{PWF})-(\ref{PDF})
There is an obvious difference between the nature of transition for $\Delta<1$, and $\Delta>1$. For the narrow distribution (first row in FIG.s \ref{PWF}, \ref{SuF}, \ref{PDF}), the system first goes into a two cluster state and then slowly these clusters get closer to be synchronized. For very small $\lambda$, (a),  there is no correlation in the system evident from FIG. \ref{PDF} (a), the frequency distribution being symmetric. Increasing $\lambda$, there exists two clusters in the system (FIG.\ref{SuF} (b)-(c)), the oscillators inside each cluster are synchronized but not in sync with the other cluster. By further increase in the coupling constant $\lambda$, the frequency distribution become asymmetric, and more oscillators from one cluster joins the other one. The phase difference shifts towards zero. It can be seen from the shift of the small cluster toward one in FIG.\ref{PDF}. (c)-(f). 

On the other hand, the wide distribution (second row in FIG. \ref{SuF}), the system stays totally out of sync up to a threshold point FIG. \ref{SuF}(k) ($\lambda=2.98$), and does an explosive transition to the synchronized state FIG. \ref{SuF}(l) ($\lambda=3$). As shown in FIG. \ref{LorentzianR} (d). Notice that close to transition FIG. \ref{PDF}(k), the distribution around zero becomes wider.

FIG.\ref{PDF} presents the probability distribution of the correlation matrixes in FIG.\ref{SuF}. At FIG.\ref{PDF} (a), the system has pick around zero. The same corresponding value $(\lambda=0.02)$ in $\Delta=2$, the distribution is narrower, meaning even less correlation. For narrow distribution, by increase in the coupling constant, small correlation rises close to one. By increasing the coupling strength $(\lambda)$, the tail at one becomes stronger as in (d) while the central pick becomes weaker with shifts towards one, corresponding to a shift in the correlation matrix toward red. There is no gradual increase in correlation in the wide distribution, but the system has a first-order transition from one state to other (k),(l).

\subsubsection{Correlation matrix, frequency distribution (backrward path)}
\begin{figure*}[!ht]
\includegraphics[scale=0.39]{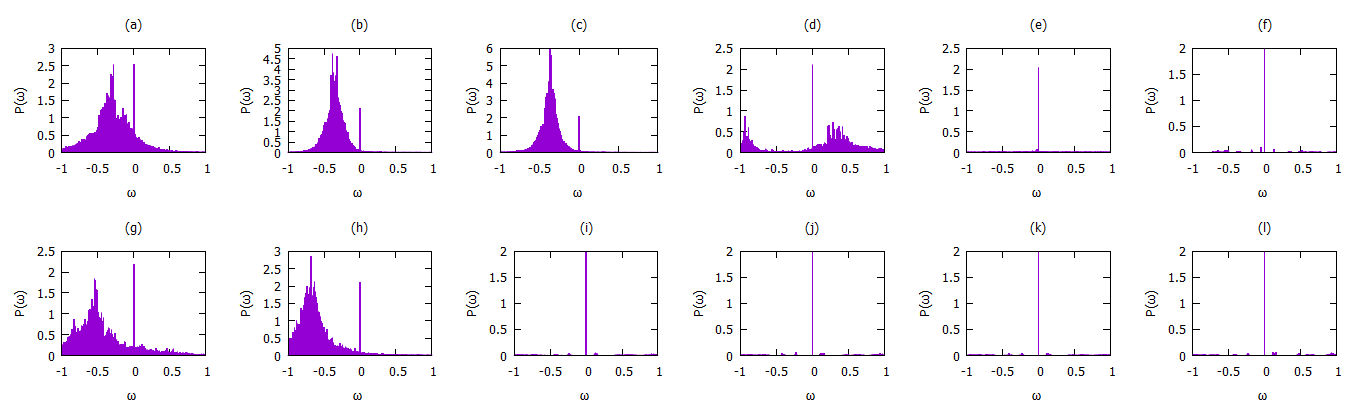}
\caption{Probability density function of the frequency (P($\omega$)) in backward transition. The first row corresponds to the case of $\Delta=0.8$, and second row for the last row $\Delta=2$. Each column corresponds to the same value of $\lambda$. In (a), (g) $\lambda=3$, in (b), (h) $\lambda=2.1$, in (c), (i) $\lambda=2.08$, n (d), (j) $\lambda=2$, in (e), (k) $\lambda=1.8$, in (f), (l) $\lambda=1$.}
\label{PWB}
\end{figure*}

\begin{figure*}[!ht]
\includegraphics[scale=0.39]{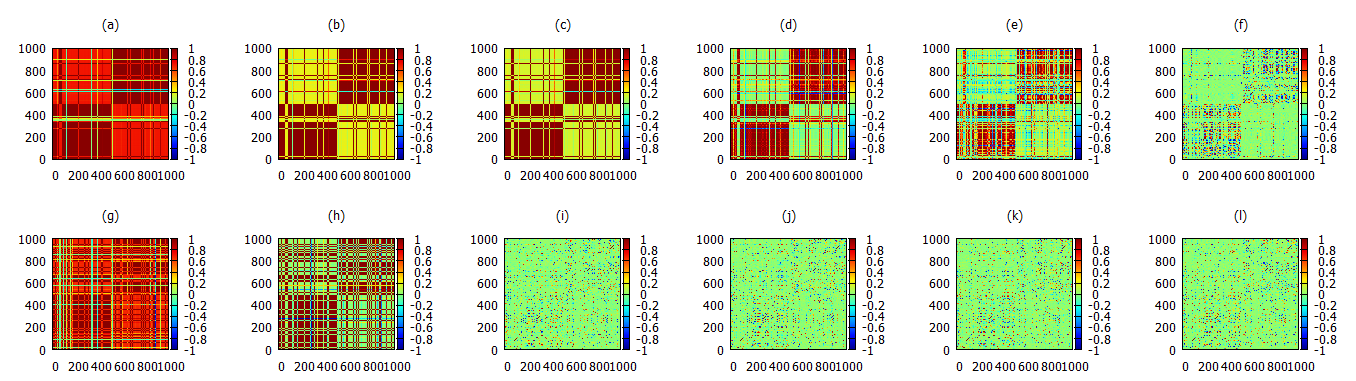}
\caption{Correlation matrix $(D_{ij})$ in backward transition. The first row corresponds to the case of $\Delta=0.8$, and second row for the last row $\Delta=2$. Each column corresponds to the same value of $\lambda$. In (a), (g) $\lambda=3$, in (b), (h) $\lambda=2.1$, in (c), (i) $\lambda=2.08$, n (d), (j) $\lambda=2$, in (e), (k) $\lambda=1.8$, in (f), (l) $\lambda=1$.}
\label{SuB}
\end{figure*}

\begin{figure*}[!ht]
\includegraphics[scale=0.39]{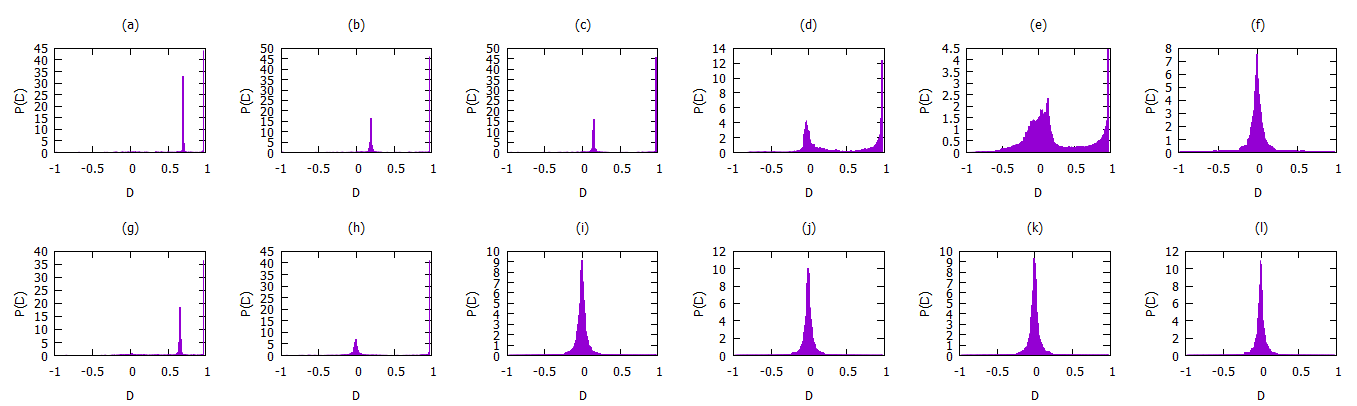}
\caption{Probability distribution of the correlation matrix P$(D_{ij})$ in backward transition. The first row corresponds to the case of $\Delta=0.8$, and second row for the last row $\Delta=2$. Each column corresponds to the same value of $\lambda$. In (a), (g) $\lambda=3$, in (b), (h) $\lambda=2.1$, in (c), (i) $\lambda=2.08$, n (d), (j) $\lambda=2$, in (e), (k) $\lambda=1.8$, in (f), (l) $\lambda=1$.}
\label{PDB}
\end{figure*}

To see the difference between the dynamic in forward and backward transition, we present the frequency distribution and correlation matrix and it's probability distribution in FIG.\ref{PWB},  \ref{SuB}, \ref{PDB} respectively. Similar to the forward plots (FIG. \ref{PWF}, \ref{PDF}, \ref{SuF}). The first row corresponds to the case of $\Delta=0.8$, and the second row $\Delta=2$. For $\Delta=0.8$, we observe three type of dynamics. For $\lambda \geq 2.08$, (a)-(c), the system is in the high order parameter state, with large population of oscillators being in frequency-locked state which shows itself as an asymmetric distribution in frequency FIG.\ref{PWB} and as a pick at one in FIG. \ref{PDB}. with a small population being out of sync. Increasing the coupling constant, (d)-(e), the probability distribution around zero gets stronger in FIG. \ref{PDB}. Increasing the coupling constant, the system goes to a full unsynchronized state (d), with a symmetric distribution around zero. In this case, the frequency distribution also has become sharp around zero, implying a symmetric distribution for the oscillators. For $\Delta=2$, on the other hand, there is no intermediate state, and the system shows a sudden transition to the unsynchronized state. At the transition point (h) to (g), there is a sudden change of the distribution from a pick at one to a distribution around zero.

\section{Summary and concluding remarks}
In conclusion, a low dimensional description of the generalized Kuramoto model has been studied analytically and checked numerically. We have demonstrated that the Ott-Antonsen ansatz is valid for the generalized Kuramoto model. 
 It gives a reasonable estimate of the separation of the dynamic in the parameter space. This means that except for the bistable region, all the attractors of the infinite-dimensional system lie in the low-dimensional manifold obtained by this ansatz. Our numeric simulations show that the transition to the synchronized state can be a two-step transition, a continuous transition, or an explosive transition with hysteresis, depending on the width of the distribution. Our result is also compatible with analytic results obtained for the exact critical coupling of the transition \cite{hu2014exact}, in case that there is hysteresis.

\section{Acknowledgments}
The authors gratefully acknowledge the Sheikh Bahaei National High-Performance Computing Center (SBNH-PCC) for providing computing facilities and time. SBNHPCC is supported by the scientific and technological department of the presidential office and Isfahan University of Technology (IUT). We acknowledge fruitful discussions with Farhad Shahbazi.

\section{Author contribution}
K.A.Samani and S.Ameli designed the research. S.Ameli performed the numerical simulation and theoretical analysis and wrote the paper with input from K.A.Samani. Both authors analysed the results, and approved the manuscript.

\bibliography{interactnlmsample}

\end{document}